\documentstyle{amsppt}
\magnification\magstep1

\NoBlackBoxes

\baselineskip=12pt

\hsize=31pc
\hoffset=4pt
\parindent=1.5em
\pageheight{45.5pc}


\define\1{\'{\i}}                           

\define\w{\omega}

\define\jp{J_+}
\define\jm{J_-}
\define\jj{J_3}

\define\co{\Delta}  

\define\pois#1#2{\left\{ {#1},{#2} \right\}}         

\define\para{\omega^2}
\define\kk{K}

\define\otra{b}
\define\e{{\text e}}
\font\head=cmbx12
\define\endhead{\rm}
\font\subhead=cmbx10
\define\endsubhead{\rm}

\define\luisH{{\Cal H}}
\define\luisP{{\Cal P}}
\define\luisM{{\Cal M}}
\define\luisK{{\Cal K}}
\define\luisD{{\Cal D}}
\define\luisC{{\Cal C}}

\font\titulo=cmbx10 scaled\magstep2


\ 
\vskip 3truecm

\centerline {\titulo  Superintegrable Deformations of}
\bigskip
\centerline {\titulo  the Smorodinsky--Winternitz Hamiltonian\footnote{Published in {\it Superintegrability in  Classical and Quantum  
Systems}, edited by P.~Tempesta, P.~Winternitz,   J.~Harnad, W.~Miller Jr., G.~Pogosyan
and M.A.~Rodr\1guez,  CRM Proceedings \& Lecture Notes, vol.~37, American Mathematical
Society, 2004 }}
\bigskip

\vskip 1truecm
\bigskip

\centerline {Angel Ballesteros$^a$, Francisco J. Herranz$^a$, Fabio Musso$^{b}$, and
Orlando Ragnisco$^{b}$}
\bigskip
\bigskip

\centerline{\it { 
${}^a$Departamento de F\1sica, Universidad de Burgos, Pza.\
Misael Ba\~nuelos s.n., }}
\centerline{\it {  E-09001 Burgos, Spain }}

\bigskip

\centerline { }
\smallskip 
\centerline{\it { 
${}^b$Dipartimento di Fisica,   Universit\`a di Roma Tre and
INFN sez.\ di Roma Tre,   }}

\centerline{\it Via Vasca
Navale 84, I-00146 Roma, Italy}
\smallskip

\bigskip
\bigskip
\bigskip
\bigskip

\noindent {\bf Abstract.}	A constructive procedure to obtain superintegrable deformations
of the classical Smorodinsky--Winternitz Hamiltonian  
by using quantum deformations of its underlying Poisson $sl(2)$ coalgebra
symmetry is introduced.  Through this example, the general connection between
coalgebra symmetry and quasi-maximal superintegrability is analysed. The notion
of comodule algebra symmetry is also shown to be applicable in order to
construct new integrable deformations of certain Smorodinsky--Winternitz
systems.

\newpage

\document


\noindent
\head 1. Introduction\endhead

\medskip
\noindent
The aim of this work is to review a (co)algebraic approach to the 
superintegrability properties of the classical Smorodinsky--Winternitz (SW)
Hamiltonian~\cite{12, 18}. As we shall see, the main consequence of making
explicit such
$sl(2)$ Poisson coalgebra symmetry is the  possibility of constructing
superintegrable deformations of the SW Hamiltonian by making use of quantum
algebra deformations of $sl(2)$. We would also like to emphasize that such
deformation procedure is rather general and can be applied to other
superintegrable Hamiltonians~\cite{7}. 

In the next Section we recall the essentials  of coalgebra
symmetry~\cite{5} and also the intrinsic superintegrability
properties of the associated Hamiltonian systems~\cite{7}. Section 3 is
devoted to the description of the coalgebra symmetry of the SW
system~\cite{1}, and a set of $(2N-2)$ functionally independent constants
of the motion (including the Hamiltonian) is deduced by making use of the
$sl(2)$ coalgebra. The non-standard deformation of
$sl(2)$~\cite{10, 16, 19} is then used (Section 4) in order to construct a
family of integrable deformations of the SW Hamiltonian with a common set of
$(2N-2)$ functionally independent deformed integrals of the motion. In
Section 5, one of these deformations is shown to be of the St\"ackel
type~\cite{17}, and a new set of $(N-1)$ integrals related with this
separability property is obtained~\cite{1}. The notion of comodule algebra
symmetry~\cite{4} is introduced in Section 6, and it is shown that some
specific SW systems have such a new type of dynamical symmetry. Once
again, this symmetry enables us to construct a new integrable (but
perhaps non-superintegrable) deformation of the SW Hamiltonian. Finally, some
remarks and open problems are briefly commented.


\bigskip
\bigskip
\noindent
\head 2. Coalgebra symmetry and superintegrability\endhead

\medskip
\noindent
We recall that a   coalgebra  $(A,\Delta)$ is a (unital, associative) algebra
$A$ endowed with a  coproduct map~\cite{9, 13}:
$$
\Delta: A
\rightarrow A\otimes A,
\tag2.1
$$ 
which is  coassociative 
$$
(\Delta \otimes id) \circ \Delta=(id \otimes \Delta) \circ \Delta ,
\tag2.2
$$
{\it i.e.}, the following diagram is a commutative one: 
$$
\CD 
A  @>{{\Delta}}>>  A\otimes A \\
@V{{\Delta}}VV  @V{{\Delta\otimes{{id}}}}VV \\ 
A\otimes A  @>{{{{id}}} \otimes \Delta} >>  A\otimes A  \otimes A  
\endCD
$$
\medskip

This ``two-fold way" for the definition of the objects on $A\otimes A\otimes A$
will be essential as far as superintegrability is concerned. Note that, in
addition, $\Delta$ has to be an algebra {homomorphism} from
$A$ to $A\otimes A$:
$$
\Delta (a\,b)=\Delta (a)\,\Delta (b) ,\qquad \forall\, a,b \in A.
\tag2.3
$$
Moreover, if $A$ is a Poisson algebra and
$$
\Delta(\pois{a}{b}_A)=\pois{\Delta(a)}{\Delta(b)}_{A\otimes A} ,\qquad \forall
a,b \in A ,
\tag2.4
$$
we shall say that $(A,\Delta)$ is a  Poisson coalgebra, which will be the
relevant object for the construction~\cite{5} of classical integrable systems
that is summarized in the sequel (see~\cite{1,2,6,14,15} for different
applications to classical and quantum systems).

Let $(A,\Delta)$  be a 
Poisson coalgebra with {generators}
$X_i$
$(i=1,\dots,l)$, 
{Casimir} function ${\Cal{C}}(X_1,\dots,X_l)$  
and {coassociative coproduct}
$\Delta\equiv\Delta^{(2)}$
which is a {Poisson map} with respect to the
Poisson bracket on $A\otimes A$ given by:
$$
\pois{X_i\otimes X_j}{X_r\otimes X_s}_{A\otimes  A}=
\{X_i, X_r\}_A\otimes X_j  X_s +
 X_i  X_r \otimes \{X_j, X_s \}_A  .
\tag2.5
$$
The {$m$-th coproduct} map $\co^{(m)}_L:A\rightarrow A\otimes A\otimes
\dots^{m)}\otimes A$
can be defined  by applying recursively the
coproduct $\co^{(2)}$ in the form
$$
\co_L^{(m)}:=(id\otimes id\otimes\dots^{m-2)}\otimes id\otimes
\co^{(2)})\circ\co_L^{(m-1)}.
\tag2.6
$$
Such an induction ensures that $\co^{(m)}_L$ is also a Poisson map.
As a consequence of the definition of $\co^{(m)}_L$, for any smooth function
${\Cal{H}}(X_1,\dots,X_l)$ we can define a $N$-sites Hamiltonian as the
$N$-th coproduct of 
${\Cal{H}}$:
$$
H^{(N)}:=\co_L^{(N)}({\Cal{H}}(X_1,\dots,X_l))=
{\Cal{H}}(\co_L^{(N)}(X_1),\dots,\co_L^{(N)}(X_l)).
\tag2.7
$$
By construction, it can be proven that the $(N-1)$ functions 
given by $(m=2,\dots,N)$
$$
C^{(m)}:= \co_L^{(m)}({\Cal{C}}(X_1,\dots,X_l))=
{\Cal{C}}(\co_L^{(m)}(X_1),\dots,\co_L^{(m)}(X_l)),
\tag2.8
$$
Poisson-commute with the Hamiltonian:
$$
\pois{C^{(m)}}{H^{(N)}}_{A\otimes
A\otimes\dots^{N)}\otimes A}=0 ,\qquad m=2,\dots,N .
\tag2.9
$$
Moreover, all these integrals of the motion are mutually in involution:
$$
\pois{C^{(m)}}{C^{(n)}}_{A\otimes
A\otimes\dots^{N)}\otimes A}=0 ,\qquad  m,n=2,\dots,N.
\tag2.10
$$

When the Hamiltonians
${\Cal{H}}$ are defined on Poisson--Lie algebras, the coproduct is ``primitive":
$ 
\co(X_i)= X_i\otimes 1+1\otimes X_i
$.
However, the Poisson analogues of quantum
algebras and groups~\cite{9, 13} are also (deformed) coalgebras
$(A_z,\Delta_z)$ (where $z$ is the deformation parameter). Consequently,
any function of the generators of a given ``quantum" Poisson algebra (with
Casimir element $C_z$) will provide, under an appropriate (deformed)
symplectic representation, an {integrable deformation} of the Hamiltonian
defined on the coalgebra
$(A,\Delta)$.


\medskip
\noindent
\subhead 2.1. Coalgebras and quasi-maximal superintegrability\endsubhead

\medskip
\noindent
Instead of (2.6), another recursion relation for
the {$m$-th coproduct} map can be defined: 
$$
\co_R^{(m)}:=(\co^{(2)}\otimes id\otimes\dots^{m-2)}\otimes
id\,)\circ\co_R^{(m-1)}.
\tag2.11
$$
Due to the coassociativity property of the coproduct, this new expression 
will provide exactly the same expressions for the $N$-th coproduct of
any generator~\cite{5}. However, if we label from
$1$ to $N$ the sites of the chain of $N$ copies of $A$, lower dimensional
coproducts
$\co^{(m)}$ (with $m<N$) will be ``different" in the sense that
$\co_L^{(m)}$ will contain objects living on the tensor product space
$ 1\otimes 2\otimes \dots\otimes m $, whilst $\co_R^{(m)}$ will be
defined on the sites
$(N-m+1)\otimes (N-m)\otimes \dots\otimes N$.
Therefore, for rank-one coalgebras the coalgebra symmetry of a given
Hamiltonian  gives rise to {two sets of $(N-1)$ integrals of the motion} that
Poisson-commute with
$H^{(N)}$~\cite{7}:

 {$\bullet$ A  set of ``left" integrals}
$\{C^{(m)}=\co_L^{(m)}({\Cal{C}}),\,m=2,\dots, N\}$:
$$
\alignedat4
& C^{(2)} &  &\equiv\Delta_L^{(2)}({\Cal C})&\quad
&\text{which is defined on the space}&\quad &1\otimes 2\\
&C^{(3)}   &  & \equiv \Delta_L^{(3)}({\Cal C})&\quad
&\text{\qquad\qquad "}&\quad &1\otimes 2\otimes 3\\
&\quad \vdots  &  &  &\quad
& &\quad &\qquad\vdots\\
&C^{(N)} &    &\equiv \Delta_L^{(N)}({\Cal C})&\quad
&\text{\qquad\qquad "}&\quad &1\otimes 2\otimes
\dots \otimes N
\endalignedat
$$

{$\bullet$ A set  of ``right"  integrals}
$\{I^{(m)}=\co_R^{(m)}({\Cal{C}}),\,m=2,\dots, N\}$:
$$
\alignedat4
& I^{(2)}  & &\equiv\Delta_R^{(2)}({\Cal C}) &\quad
&\text{ on the space}&\quad (N-1)\otimes N\\
&I^{(3)} & &\equiv\Delta_R^{(3)}({\Cal C})&\quad
&\text{\qquad\qquad "}&\quad(N-2)\otimes(N-1)\otimes N\\
&\quad\vdots & &   &\quad
&\text{\qquad\qquad  }&\quad     \vdots\qquad\qquad \\
&I^{(N)} & &\equiv\Delta_R^{(N)}({\Cal C}) &\quad
&\text{\qquad\qquad "}&\quad  1\otimes
\dots\otimes (N-2)\otimes(N-1)\otimes N
\endalignedat
$$

Note that $ C^{(N)}\equiv I^{(N)}$. Thus, if all  these integrals are
functionally independent, we obtain an explicit construction of
{``quasi-maximally superintegrable systems"}, since the coalgebra generates a
set of $(2N -2)$ functions in involution
$$
\{H^{(N)},C^{(2)},\dots,C^{(N-1)},C^{(N)}\equiv I^{(N)},I^{(N-1)}, 
\dots,I^{(2)}\}.
\tag2.12
$$

We remark that, in some cases, one more independent integral could exist
(leading to a maximally  superintegrable system), but such
remaining constant of the motion cannot be deduced from the coalgebra
symmetry.

\newpage


\medskip
\medskip
\medskip
\noindent
\head 3. Coalgebra symmetry of the SW Hamiltonian\endhead

\medskip
\noindent
Let us consider the $sl(2)$ Poisson coalgebra~\cite{1, 16}: 
$$
\aligned
&\{\jj,\jp\}=2\jp ,\qquad 
\{\jj,\jm\}=-2\jm ,\qquad \{\jm,\jp\}=4\jj ,\\
&\co(J_i)=1\otimes J_i + J_i\otimes 1,\qquad i=+,-,3,
\endaligned
\tag3.1
$$
with  Casimir function  ${\Cal C}=\jj^2 - \jm \jp$.
A one-particle {symplectic realization} of this coalgebra is given by 
$$ 
D(\jm)=q_1^2 ,\qquad
D(\jp)=p_1^2 +
\frac{\otra_1}{q_1^2} ,\qquad
D(\jj)= q_1 p_1  ,
\tag3.2
$$
where $\pois{q_1}{p_1}=1$. Note that, under this realization, $
D({\Cal C})=-\otra_1$.

If we consider the following {Hamiltonian function}:
$$
{\Cal H}= \jp+\para  \jm ,
\tag3.3
$$
its one-particle realization is just
$$
H^{(1)}=D ({\Cal H})=p_1^2 + \w^2 q_1^2 +
\frac{\otra_1}{q_1^2}.
\tag3.4
$$

The {$2$-particle realization} of the coalgebra is obtained  through the
coproduct:
$$
\aligned
&  (D\otimes D)(\Delta^{(2)}({\jm}))=f_-^{(2)}= 
q_1^2 + q_2^2 ,\\
&  (D\otimes D)(\Delta^{(2)}({\jp}))=f_+^{(2)} = 
p_1^2 + p_2^2 + \frac{\otra_1}{q_1^2}+ \frac{\otra_2}{q_2^2} ,\\
&  (D\otimes D)(\Delta^{(2)}({\jj}))=f_3^{(2)}= q_1 p_1 + q_2 p_2 .
\endaligned
\tag3.5
$$
Hence the associated {$2$-particle} Hamiltonian is
$$ 
H^{(2)}=(D\otimes D)(\Delta^{(2)}({\Cal H}))= \sum_{i=1}^2 \biggl( p_i^2 +\para
q_i^2+
\frac{\otra_i}{q_i^2}\biggr)  ,
\tag3.6
$$
which is just the $N=2$ SW Hamiltonian.
A (both left and right) {constant of the motion} for $H^{(2)}$ is given by the
coproduct of the Casimir:
$$  
 C^{(2)}=(D\otimes D)(\Delta^{(2)}({\Cal C}))=  - ({q_1}{p_2} - {q_2}{p_1})^2
- \left(
\otra_1\frac{q_2^2}{q_1^2}+\otra_2\frac{q_1^2}{q_2^2}\right)
-\sum_{i=1}^2 \otra_i .
\tag3.7
$$ 

In general, the  {$N$-particle realization} is obtained by applying the
$\Delta^{(N)}$ map:
$$
\aligned
& (D\otimes D\otimes\dots^{N)}\otimes D)(\Delta_L^{(N)}({\jm}))=f_-^{(N)}= 
\sum_{i=1}^N   q_i^2 ,\\
& (D\otimes D\otimes\dots^{N)}\otimes D)(\Delta_L^{(N)}({\jp}))=f_+^{(N)}
=\sum_{i=1}^N  
\biggl(p_i^2 + \frac{\otra_i}{q_i^2}\biggr) ,\\
&  (D\otimes D\otimes\dots^{N)}\otimes D)(\Delta_L^{(N)}({\jj}))=f_3^{(N)}=
\sum_{i=1}^N q_i p_i  ,
\endaligned
\tag3.8
$$
and the 
 {$N$-particle} Hamiltonian given by the coalgebra is just the SW system:
$$ 
H^{(N)}=(D\otimes D\otimes\dots^{N)}\otimes D)(\Delta_L^{(N)}({\Cal
H}))=\sum_{i=1}^N \biggl( p_i^2 +\para q_i^2+
\frac{\otra_i}{q_i^2}\biggr) . 
\tag3.9
$$
The first set of $(N-1)$ (left) {constants of the motion} in involution
turns out to be $(m=2,\dots,N)$:
$$  
 C^{(m)}=(D\otimes D\otimes\dots^{m)}\otimes D)(\Delta_L^{(m)}({\Cal C}))
=-\sum_{i<j}^m{I_{ij}} -\sum_{i=1}^m \otra_i ,
\tag3.10
$$
where
$$  
I_{ij}=({q_i}{p_j} - {q_j}{p_i})^2 + \left(
\otra_i\frac{q_j^2}{q_i^2}+\otra_j\frac{q_i^2}{q_j^2}\right).
\tag3.11
$$
In this way, the complete integrability of the SW Hamiltonian is extracted from
the coalgebra symmetry of the model~\cite{1}.


\bigskip

\medskip
\noindent
\subhead 3.1. Coalgebraic superintegrability\endsubhead

\medskip
\noindent
Further to the integrability, the coalgebra symmetry also underlies the
superintegrability of the SW Hamiltonian since besides the ``left
integrals" $C^{(m)}$ (3.10), there exists a set of  ``right" ones $I^{(m)}$
given by $(m=2,\dots,N)$:
$$ 
 I^{(m)} = (D\otimes D\otimes\dots^{m)}\otimes D)(\Delta_R^{(m)}({\Cal C})) 
 =  -\!\!\!\!\!\sum_{N-m+1\leq i<j}^N{
\!\!\!\!\! I_{ij}} - \!\!\! \sum_{i=N-m+1}^N \!\!\!\otra_i  .
\tag3.12
$$ 

The {functional independence} of all these integrals follows from the
properties of their $I_{ij}$ building blocks. Let us firstly consider the
  $N=3$ integrals:
$$
\aligned
&  C^{(2)}=- {I_{12}} -( \otra_1+ \otra_2)  ,\qquad I^{(2)}=- {I_{23}} -(
\otra_2+ \otra_3) , \\
&  C^{(3)}\equiv  I^{(3)}=- {I_{12}} - {I_{13}} - {I_{23}}
-( \otra_1+ \otra_2+
\otra_3) , 
\endaligned
$$
which are functionally independent, since $C^{(3)}$ contains the 
${I_{13}}$ term. Similarly for the $N=4$ case, where the integrals
coming from the coalgebra read
$$
\aligned
& C^{(2)}=- {I_{12}} -( \otra_1+ \otra_2) ,\qquad I^{(2)}=- {I_{34}} -( \otra_3+  
\otra_4) ,\\
&  C^{(3)}=- {I_{12}} - {I_{13}} - {I_{23}} -( \otra_1+ \otra_2+
\otra_3) ,\\& I^{(3)}=- {I_{23}} - {I_{24}} - {I_{34}}
-( \otra_2+ \otra_3+
\otra_4)    ,\\
&  C^{(4)} \equiv I^{(4)} =- {I_{12}} - {I_{13}} - {I_{14}} - {I_{23}}- {I_{24}}
- {I_{34}} -( \otra_1+ \otra_2+
\otra_3+\otra_4)  .
\endaligned
$$
Once again, the fact that ${I_{14}}$  does appear within $C^{(4)}$ implies the
functional independence  of the full set of integrals. In the $N$-dimensional
 case, by following the same construction, the functional independence is proven
by considering that   $C^{(N)}\equiv I^{(N)}$  is the only integral that
contains the ${I_{1N}}$ term. Since the $N$-dimensional SW Hamiltonian is, by
construction, functionally independent of the $C^{(N)}$ integral, the
quasi-maximal superintegrability of the SW Hamiltonian is proven. Finally, in
this particular (separable) case we can take any of the one-particle SW
Hamiltonians as the remaining independent integral leading to the full maximal
superintegrability of the system.

Furthermore, we stress that a  {much more general family of coalgebra-symmetric
quasi-maximally superintegrable Hamiltonians} than (3.9) can also be 
defined~\cite{1}. For instance, let us consider the Hamiltonian function
$$
{\Cal H}=\jp+ {\Cal F}(\jm) ,
\tag3.13
$$
where ${\Cal F}(\jm)$ is an {arbitrary smooth  function} of $\jm$.
By construction, any $N$-particle Hamiltonian of the form
$$
 H^{(N)}
=    f_+^{(N)} +{\Cal F}(f_-^{(N)}) 
=\sum_{i=1}^N \biggl( p_i^2 + \frac{\otra_i}{q_i^2}  \biggr)+
{\Cal F}\biggl(\sum_{i=1}^N q_i^2\biggr) ,
\tag3.14
$$
is  completely integrable  (moreover, quasi-maximally superintegrable), and
its constants of the motion are the previous sets $C^{(m)}$ and $I^{(m)}$ .
Note that in the case $N=3$, this system is just one of the   superintegrable
potentials given by Evans~\cite{11}.


\bigskip
\bigskip
\noindent
\head 4. A superintegrable deformation of   the SW Hamiltonian\endhead

\medskip
\noindent
Now we   consider the Poisson analogue~\cite{1} of the ``non-standard" 
deformation of
$sl_z(2)$~\cite{16}: 
$$
\aligned
& \{\jj,\jp\}=2 \jp \cosh z\jm  ,\\
& \{\jj,\jm\}=-2\,\frac {\sinh z\jm}{z} ,\qquad
 \{\jm,\jp\}=4 \jj  . 
\endaligned
\tag4.1
$$
A deformed Casimir  function for $sl_z(2)$ is found to be:
$$
{\Cal C}_z=\jj^2  - \frac {\sinh z\jm}{z} \jp .
\tag4.2
$$
The deformed coproduct map   
$\Delta_z:sl_z(2)\rightarrow sl_z(2)\otimes sl_z(2)$ is given by:
$$
\aligned
 &\Delta_z(\jm)=  \jm \otimes 1+
1\otimes \jm ,\\
 &\Delta_z(\jp)=\jp \otimes \e^{z \jm} + \e^{-z \jm} \otimes
\jp ,\\
 &\Delta_z(\jj)=\jj \otimes \e^{z \jm} + \e^{-z \jm} \otimes \jj .
\endaligned
\tag4.3
$$

Let us mimic the construction performed in Section 3 by taking again the
function (3.3) for  ${\Cal H}$.
A one-particle deformed symplectic realization of $sl_z(2)$ is:
$$
\aligned
&D_z(\jm)=q_1^2, \qquad
 D_z(\jp)=\frac {\sinh z q_1^2}{z q_1^2}\, p_1^2 +
\frac{z \otra_1}{\sinh z q_1^2} ,\\
 &D_z(\jj)=\frac {\sinh z q_1^2}{z q_1^2}\, q_1 p_1  ,
\endaligned
\tag4.4
$$
which is characterized by the Casimir function $C_z^{(1)}=D_z({\Cal
C}_z)=-\otra_1$. The associated {one-particle Hamiltonian} is just:
$$
H_z^{(1)}=D_z({\Cal H})=\frac {\sinh z q_1^2}{z q_1^2} \,p_1^2 +
\frac{z \otra_1}{\sinh z q_1^2} + \para \,q_1^2 ,
\tag4.5
$$
and the {$2$-particle symplectic realization} is obtained through $\Delta_z$:
$$
\aligned
& (D_z\otimes D_z)(\Delta_z^{(2)}({\jm}))={\tilde f}_-^{(2)}= 
q_1^2 + q_2^2 ,\\
 & (D_z\otimes D_z)(\Delta_z^{(2)}({\jp}))={\tilde f}_+^{(2)} = 
\left( \frac {\sinh z q_1^2}{z q_1^2} \, p_1^2  +\frac{z \otra_1}{\sinh z
q_1^2} \right)   \e^{  z q_2^2 }\\
&\qquad\qquad \qquad\qquad \qquad\qquad \qquad +
\left( \frac {\sinh z q_2^2}{z q_2^2} \, p_2^2  +\frac{z \otra_2}{\sinh z
q_2^2} \right)   \e^{ - z q_1^2 } ,\\
 & (D_z\otimes D_z)(\Delta_z^{(2)}({\jj}))={\tilde f}_3^{(2)}=  
\frac {\sinh z q_1^2}{z q_1^2}\,  q_1p_1 \, \e^{  z q_2^2 }+
\frac {\sinh z q_2^2}{z q_2^2}\,  q_2p_2 \, \e^{ - z q_1^2 } .
\endaligned
\tag4.6
$$
As a consequence, a deformation of the 
 {$2$-particle} SW Hamiltonian is obtained as 
$  H_z^{(2)}=(D_z\otimes D_z)(\Delta^{(2)}({\Cal H}))$. Namely,
$$
\aligned
&  H^{(2)}_z = 
\left( \frac {\sinh z q_1^2}{z q_1^2} \, p_1^2  +\frac{z \otra_1}{\sinh z
q_1^2} \right)   \e^{  z q_2^2 } \\
&\qquad\qquad  +
\left( \frac {\sinh z q_2^2}{z q_2^2} \, p_2^2  +\frac{z \otra_2}{\sinh z
q_2^2} \right)   \e^{ - z q_1^2 }   + \para(
q_1^2+q_2^2) .
\endaligned
\tag4.7
$$
Note that separability  is destroyed under  deformation. The corresponding
{constant of the motion} is 
$ C_z^{(2)}=(D_z\otimes D_z)(\Delta^{(2)}({\Cal C}_z))$. Explicitly, 
$$
\aligned
&   C^{(2)}_z=- 
\frac {\sinh z q_1^2}{z  q_1^2}\,
\frac {\sinh z q_2^2}{z q_2^2}
\left({q_1}{p_2} - {q_2}{p_1}\right)^2  \e^{ z
(q_2^2 - q_1^2)} -  \otra_1  \e^{2 z q_2^2 }-  \otra_2  \e^{-2 z q_1^2 }  \\
& \qquad\qquad - 
\left( \otra_1\, \frac {\sinh z q_2^2}{\sinh z q_1^2}
+ \otra_2\, \frac {\sinh z q_1^2}{\sinh z q_2^2} \right)  \e^{ z
(q_2^2 - q_1^2)} .
\endaligned
\tag4.8
$$

The generic {$m$-particle symplectic realization} is then obtained through the
(either right or left) $m$-th deformed coproduct and reads:
$$
\aligned
& (D_z\otimes \dots^{m)}\otimes D_z)(\Delta_{z,L}^{(m)}({\jj}))={\tilde
f}_-^{(m)}= \sum_{i=1}^m q_i^2 ,\\
&  (D_z\otimes \dots^{m)}\otimes D_z)(\Delta_{z,L}^{(m)}({\jp}))\\
&\qquad\qquad\qquad \quad  ={\tilde
f}_+^{(m)} =\sum_{i=1}^m
\left( \frac {\sinh z q_i^2}{z q_i^2}  p_i^2  +\frac{z \otra_i}{\sinh z
q_i^2} \right)  \exp\left\{z \kk_i^{(m)}(q^2) \right\} ,\\
&(D_z\otimes \dots^{m)}\otimes D_z)(\Delta_{z,L}^{(m)}({\jm}))={\tilde
f}_3^{(m)}=\sum_{i=1}^m
\frac {\sinh z q_i^2}{z q_i^2}  q_ip_i  \exp\left\{z \kk_i^{(m)}(q^2) \right\} 
,
\endaligned
\tag4.9
$$
where the ``long-range" interaction is encoded within the functions
$$
\aligned
\kk_i^{(m)}(q^2)& =  -
\sum_{k=1}^{i-1}  q^2_k+ 
\sum_{l=i+1}^m   q^2_l ,\\
 \kk_{ij}^{(m)}(q^2) &= \kk_i^{(m)}(q^2) \!+\! \kk_j^{(m)}(q^2)\\
 & =    - 2\sum_{k=1}^{i-1}   q^2_k  -    q^2_i  +    q^2_j  + 
2\sum_{l=j+1}^m   q^2_l\, ,\quad (i<j)  .
\endaligned
\tag4.10
$$

In this way, the {$N$-particle} deformed SW Hamiltonian is defined as:
$$
\aligned
 H^{(N)}_z &=  {\tilde f}_+^{(N)}+\para {\tilde f}_-^{(N)}\\
&=\sum_{i=1}^N
\left( \frac {\sinh z q_i^2}{z q_i^2} \, p_i^2  +\frac{z \otra_i}{\sinh z
q_i^2} \right)  \exp\left\{z \kk_i^{(N)}(q^2) \right\} + \para\,\sum_{i=1}^N
q_i^2 .
\endaligned
\tag4.11
$$
And the (left) constants of the motion in involution with $H^{(N)}_z$ are:
$$
\aligned
&   C^{(m)}_z=- \sum_{i<j}^m 
\frac {\sinh z q_i^2}{z  q_i^2}\,
\frac {\sinh z q_j^2}{z q_j^2}
\left({q_i}{p_j} - {q_j}{p_i}\right)^2  \exp\left\{ z
\kk_{ij}^{(m)}(q^2) \right\} \\
&\qquad\qquad  - \sum_{i<j}^m 
\left( \otra_i\, \frac {\sinh z q_j^2}{\sinh z q_i^2}
+ \otra_j \, \frac {\sinh z q_i^2}{\sinh z q_j^2} \right)  \exp\left\{ z
\kk_{ij}^{(m)}(q^2)\right\} \\
 & \qquad\qquad 
- \sum_{i=1}^m \otra_i  \exp\left\{2 z
\kk_{i}^{(m)}(q^2)\right\} .
\endaligned
\tag4.12
$$
These deformed integrals can  also be written as
$$  
C_z^{(m)}= -\sum_{i<j}^m{I_{ij}^z}\,\exp\left\{ z
\kk_{ij}^{(m)}(q^2) \right\} -\sum_{i=1}^m \otra_i  \exp\left\{2 z
\kk_{i}^{(m)}(q^2) \right\} ,
\tag4.13
$$ 
where we have defined the following analogues of the $I_{ij}$ symbols (3.11):
$$  
I_{ij}^z= \frac {\sinh z q_i^2}{z  q_i^2}\,
\frac {\sinh z q_j^2}{z q_j^2}
\left({q_i}{p_j} - {q_j}{p_i}\right)^2  
 +\left( \otra_i\, \frac {\sinh z q_j^2}{\sinh z q_i^2}
+ \otra_j \,\frac {\sinh z q_i^2}{\sinh z q_j^2} \right) .
\tag4.14
$$


\medskip
\medskip
\noindent
\subhead 4.1. Coalgebraic superintegrability of the deformation\endsubhead

\medskip
\noindent
By following the very same procedure as in the non-deformed case,
the set of deformed right integrals
$I^{(m)}_z$ can be easily constructed and reads  
$$  
 I^{(m)}_z= -\!\!\!\!\sum_{N-m+1\leq i<j}^N \!\!\!\!
I_{ij}^z \exp\left\{ zR_{ij}^{(m)}(q^2)  \right\} 
-\!\!\!\! \sum_{i=N-m+1}^N \!\!\!\! \otra_i  \exp\left\{2 z
R_{i}^{(m)}(q^2)   \right\} ,
\tag4.15
$$ 
where the ``long-range" interaction $R$-functions, similar to (4.10), are
defined by
$$
\aligned
R_i^{(m)}(q^2) & =  - \!\!\sum_{p=N-m+1}^{i-1}\!\!  q^2_p+ 
\sum_{l=i+1}^N   q^2_l ,\\
 R_{ij}^{(m)}(q^2)&  =R_i^{(m)}(q^2)+R_j^{(m)}(q^2)   \\
&=   - 2 \!\! \sum_{p=N-m+1}^{i-1}  \!\!
q^2_p  -    q^2_i  +    q^2_j  +  2\sum_{l=j+1}^N   q^2_l , \quad (i<j) .
\endaligned
\tag4.16
$$

  The {functional independence} of the left and right deformed integrals
follows from the fact that they are {analytic} in the deformation parameter $z$:
$$  
 C_z^{(m)}=  C^{(m)} + o[z] ,\qquad  I_z^{(m)}=  I^{(m)} + o[z] .
$$ 
As they are functionally independent at $z=0$, they they will be so in the whole complex $z$-plane,
up to isolated points.

Let us explicitly write such integrals in the 
$N=3$ case:
$$
\aligned
& C^{(2)}_z=- {I_{12}^z} \exp\left\{ z
\kk_{12}^{(2)}(q^2) \right\} -\sum_{i=1}^2 \otra_i  \exp\left\{2 z
\kk_{i}^{(2)}(q^2) \right\} ,\\ 
&  C^{(3)}_z\equiv I^{(3)}_z=- {I_{12}^z} \exp\left\{ z
\kk_{12}^{(3)}(q^2) \right\} - {I_{13}^z} \exp\left\{ z
\kk_{13}^{(3)}(q^2) \right\}\\
&\qquad\qquad \qquad  - {I_{23}^z}\exp\left\{ z
\kk_{23}^{(3)}(q^2) \right\} -\sum_{i=1}^3 \otra_i \exp\left\{2 z
\kk_{i}^{(3)}(q^2) \right\} \\
&   I^{(2)}_z=- {I_{23}^z}\exp\left\{z
R_{23}^{(2)}(q^2) \right\} -\sum_{i=2}^3 \otra_i \exp\left\{2 z
R_{i}^{(2)}(q^2) \right\} ,
\endaligned
$$
where the $K$ and $R$-functions  involved in the previous expressions read
$$
\alignedat2
&\kk_{1}^{(2)}(q^2)=  q_2^2 ,&\qquad
& \kk_{2}^{(2)}(q^2)=  - q_1^2 ,\qquad\quad
\kk_{12}^{(2)}(q^2)=  - q_1^2+ q_2^2 ,\\
&\kk_{1}^{(3)}(q^2)=  q_2^2 + q_3^2 ,&\qquad
&\kk_{12}^{(3)}(q^2)=   - q_1^2 + q_2^2 + 2 q_3^2   ,\\
&\kk_{2}^{(3)}(q^2)=  -q_1^2 + q_3^2 ,&\qquad
&\kk_{13}^{(3)}(q^2)=    - q_1^2 +  q_3^2   ,\\
&\kk_{3}^{(3)}(q^2)=  - q_1^2 - q_2^2 ,&\qquad &
\kk_{23}^{(3)}(q^2)=   - 2 q_1^2 - q_2^2 + q_3^2 ,\\
&R_{2}^{(2)}(q^2)=   q_3^2 ,&\qquad
&R_{3}^{(2)}(q^2)=  - q_2^2 ,\qquad\quad
R_{23}^{(2)}(q^2)=  -q_2^2 + q_3^2. 
\endalignedat
$$

Moreover, the following family of  $N$-dimensional Hamiltonian systems  is also
quasi-maximally superintegrable:
$$
{\Cal H}=\jp+ {\Cal F}(\jm) ,
\tag4.17
$$
where ${\Cal F}(\jm)$ is an  arbitrary smooth  function  of $\jm$.
Explictly, 
$$ 
\aligned
H_z^{(N)} 
&={\tilde f}_+^{(N)}+{\Cal F}({\tilde f}_-^{(N)})\\
&=\sum_{i=1}^N
\left( \frac {\sinh z q_i^2}{z q_i^2}  p_i^2  +\frac{z \otra_i}{\sinh z
q_i^2} \right)  \exp\left\{z \kk_i^{(N)}(q^2) \right\}
 + {\Cal F}\left(\sum_{i=1}^N q_i^2\right) ,
\endaligned
\tag4.18
$$
will Poisson-commute with all the $C^{(m)}_z$ and $I^{(m)}_z$.


\bigskip
\bigskip
\noindent
\head 5. A deformation of St\"ackel type\endhead

\medskip
\noindent
It is obvious that the SW Hamiltonian (3.9) is a Liouville system, and another
possible set of integrals of  motion in involution is given by
$$
M_i=p_i^2 +\para q_i^2+
\frac{\otra_i}{q_i^2}- \frac {H^{(N)}}{N} ,\qquad
i=1,\dots,N ,
\tag5.1
$$
where $\sum_{i=1}^N M_i=0$.
In order to get the 
{maximal superintegrability} of the non-deformed SW Hamiltonian
we can take any  of these integrals in order to complete, in a functionally
independent way, the $C^{(m)}$ and
$I^{(m)}$ sets of constants of the motion. On the contrary, in the deformed case
(4.11), the separability is broken due to the long-range interaction introduced
by the deformation. However,  since the coalgebra construction allows for
an  infinite family of  deformed Hamiltonians, and all of
them Poisson-commute with the same $C^{(m)}$ and $I^{(m)}$ sets,  it could
happen that another choice of the dynamical Hamiltonian could fulfil  the
separability conditions.

This is the case if we consider the {Hamiltonian} function~\cite{1}:
$$
{\Cal H}= \jp \e^{ z \jm} +\para  \left(\frac{\e^{2 z
\jm}-1}{2 z}\right) .
\tag5.2
$$
By introducing the $N$-th particle symplectic realization (4.9) we obtain
$$
\aligned
&H^{(N)}_z =  \sum_{i=1}^N \frac {\sinh z q_i^2}{z q_i^2}\, 
\e^{z q_i^2}\exp\left\{ {2z \sum_{k=i+1}^N q_k^2} \right\}
\left( p_i^2  +  \otra_i \left(\frac{z q_i}{\sinh z
q_i^2}\right)^2  \right)  \\
 &\qquad\qquad 
+\para  \left(\frac{\exp\left\{2 z\sum_{j=1}^N q_j^2\right\}-1}{2
z}\right) ,
\endaligned
\tag5.3
$$
which has the form of a   St\"ackel system 
$$
H^{(N)}_z=\sum_{i=1}^N a_i(q_1,\dots,q_N) \left(\frac 12 p_i^2 +
U_i(q_i)\right) ,
\tag5.4
$$
provided that 
$$
\aligned
&a_i(q_1,\dots,q_N)=
2\,\frac {\sinh z q_i^2}{z q_i^2}\,
\e^{z q_i^2}\exp\left\{ {2z \sum_{k=i+1}^N q_k^2}\right\} ,\qquad
i=1,\dots,N ,\\
& U_1(q_1)=
\frac {\otra_1}{2} \left(\frac{z q_1}{\sinh z
q_1^2}\right)^2 +\frac{\w^2}{4z}\, \e^{z q_1^2}
\,\frac {z q_1^2}{\sinh z q_1^2} ,\\
 &U_i(q_i)=\frac {\otra_i}{2} \left(\frac{z q_i}{\sinh
z q_i^2}\right)^2 ,\qquad i=2,\dots,N-1 ,\\
&U_N(q_N)=
\frac {\otra_N}{2} \left(\frac{z q_N}{\sinh z
q_N^2}\right)^2 -\frac{\w^2}{4z}\, \e^{-z q_N^2}
\,\frac {z q_N^2}{\sinh z q_N^2} .
\endaligned
\tag5.5
$$

St\"ackel's theorem claims that a
    Hamiltonian (5.4) admits separation
of variables in the   Hamilton--Jacobi equation  if and
only if there exists an $N\times N$ matrix $B$ with entries
$b_{ij}(q_j)$  such that
$$
\text{det}\,B\ne 0 ,\qquad \sum_{j=1}^N b_{ij}(q_j)
a_j(q_1,\dots,q_N)=\delta_{i1} .
\tag5.6
$$
And this is the case for the new deformed Hamiltonian.
The non-vanishing entries of $B$ and its determinant are
found to be
$$
\aligned
&b_{1N}(q_N)=\frac {z q_N^2}{2\sinh z q_N^2}\,\e^{-z q_N^2} ,
\qquad
b_{i\,i-1}(q_{i-1})=\frac {z q_{i-1}^2}{\sinh z q_{i-1}^2}\,\e^{-z
q_{i-1}^2} ,\\
& b_{ii}(q_{i})=-\frac {z q_{i}^2}{\sinh z q_{i}^2}\,\e^{z q_{i}^2} ,
\qquad
i=2,\dots,N , \\
&\text{det}\,B =\frac 12 \prod_{i=1}^N
\frac {z q_{i}^2}{\sinh z q_{i}^2}\,\e^{-z q_{i}^2}.
\endaligned
\tag5.7
$$

As a consequence, 
St\"ackel's theorem gives us a new set of $N$ functionally independent
integrals of motion  in involution
$$
Z_j=\sum_{i=1}^N a_{ij} \left(\frac 12 p_i^2 +
U_i(q_i)\right) ,\qquad j=1,\dots,N ,
\tag5.8
$$
where $a_{ij}$ are the entries  of $B^{-1}$. Then
$a_{i1}=a_i$, so that the first integral $I_1$ is just the
Hamiltonian. In our case, the non-zero functions
$a_{ij}$ turn out to be
$$
\aligned
& a_{i1}=2\,\frac {\sinh z q_{i}^2}{z q_{i}^2}\,\e^{z q_{i}^2}
\exp\left\{ {2z \sum_{k=i+1}^N q_k^2}\right\} ,\qquad i=1,\dots,N,\\
& a_{ij}= \frac {\sinh z q_{i}^2}{z q_{i}^2}\,\e^{z q_{i}^2}
\exp\left\{ {2z \sum_{k=i+1}^{j-1} q_k^2}\right\},\qquad i=1,\dots,N,
\quad i<j .
\endaligned
\tag5.9
$$

The new set of $N-1$ conserved quantities is given by ($j=2,\dots,N$):
$$
\aligned
& Z_j^z=\sum_{i=1}^{j-1} 
\frac {\sinh z q_{i}^2}{2z q_{i}^2}\,\e^{z q_{i}^2}
\exp\left\{ {2z \sum_{k=i+1}^{j-1} q_k^2}\right\}
\left(p_i^2
+  {\otra_i} \left(\frac{z q_i}{\sinh
z q_i^2}\right)^2\right)\cr
&  \qquad\qquad +
 \frac{\w^2}{4z} \bigg( \exp\left\{ {2z \sum_{k=1}^{j-1} q_k^2}\right\} -1
\bigg) .
\endaligned
\tag5.10
$$
Thus, for instance, the  function $Z^z_2$ can be taken as the remaining constant
of the motion, which  together with the family of ``left" and ``right" ones 
prove  the maximal superintegrability of the Hamiltonian (5.3).

 In the $z\rightarrow 0$ limit, the  integrals (5.10) reduce to
$$
  Z_j^0=\frac12 \sum_{i=1}^{j-1} 
\left(p_i^2+  \frac{\otra_i}{q_i^2}\right)+\frac12
\w^2 \sum_{k=1}^{j-1} q_k^2 .
\tag5.11
$$


\bigskip
\bigskip
\noindent
\head 6. Comodule algebra symmetry\endhead

\medskip
\noindent
 The notion of coproduct can be generalized by introducing the so called
``coactions"~\cite{13}. A (right)  coaction  of a Hopf algebra $(A,\Delta)$ 
on a vector  space $V$  is a map $\phi: V
\rightarrow V\otimes A$ such that 
$$
(\phi \otimes id) \circ \phi=(id \otimes \Delta) \circ \phi ,
\tag6.1
$$
that is, if the following diagram is commutative: 
$$
\CD 
V  @>{{\phi}}>>  V\otimes A \\
@V{{\phi}}VV  @V{{\phi\otimes{{id}}}}VV \\ 
V\otimes A  @>{{{{id}}} \otimes \Delta} >>  V\otimes A  \otimes A  
\endCD
$$
\medskip

If $V$ is an algebra, we shall say that $V$ is an  $A$-comodule
algebra  if the coaction
$\phi$ is a homomorphism on $V$
$$
\phi(a\,b)=\phi(a)\,\phi(b) ,\qquad \forall a,b \in V.
\tag6.2
$$
Moreover, if $V$ is endowed with a Poisson structure and $A$ is a Poisson-Hopf
algebra, $V$ will also be a   Poisson $A$-comodule algebra  if:
$$
 \phi(\pois{a}{b}_V)=\pois{\phi(a)}{\phi(b)}_{V\otimes A} ,\qquad \forall a,b
\in V .
\tag6.3
$$

Note that any Hopf algebra $A$ is an $A$-comodule
algebra with respect to $A$ provided that  $\phi\equiv \Delta$. 
The construction of integrable systems by making use of comodule algebras has
been recently introduced in~\cite{4} by defining recursively the 
$N$-th coaction as a homomorphism that maps  $V$ within $V\otimes
{A\otimes\cdots^{(N-1)}\otimes A}$. Let
$\{X_1,\dots,X_l\}$ be the generators of $V$ and let ${C}$ be a
Casimir function/operator of $V$. It can be proven~\cite{4} that
 the Hamiltonian
$$
H^{(N)}:=\phi^{(N)}({\Cal{H}}(X_1,\dots,X_l))=
{\Cal{H}}(\phi^{(N)}(X_1),\dots,\phi^{(N)}(X_l)),
\tag6.4
$$
together with the following  iterated (left) coactions of the Casimir are a set
of $N$ functions in involution and functionally independent $(m=2,\dots,N)$:
$$
C^{(m)}:= \phi^{(m)}(C(X_1,\dots,X_l))=
C(\phi^{(m)}(X_1),\dots,\phi^{(m)}(X_l)) .
\tag6.5
$$
We stress that the ``right" integrals cannot be defined in this approach, thus
the superintegrability of comodule symmetric systems cannot  be ensured
algebraically, and it has to be analysed in each particular case.


\medskip
\bigskip
\noindent
\subhead 6.1. Comodule algebra symmetry of the SW Hamiltonian\endsubhead

 \medskip
\noindent
Let us now describe an integrable deformation of the $N=2$ SW Hamiltonian
(3.6) with comodule algebra symmetry.  We take as the {Poisson-Hopf algebra $A$}
the following Poisson version of a non-standard deformation~\cite{3} of the
Schr\"odinger algebra ${h_6^\sigma}$~\cite{8}:
$$
\alignedat4
& \{\luisD,\luisP\}=-\luisP,& \qquad   &\{\luisD,\luisK\}=\luisK,& \qquad
&\{\luisK,\luisP\}= \luisM ,&\qquad &\{\luisM,\,\cdot\,\}=0,\\
& \{\luisD,\luisH\}=-2\luisH ,&\qquad  &\{\luisD,\luisC\}=2\luisC,& \qquad
&\{\luisH,\luisC\}=\luisD ,& \qquad &\{\luisH,\luisP\}=0,\\
& \{\luisP,\luisC\}= - \luisK,&  \qquad  &
 \{\luisK,\luisH\}= \luisP,&\qquad  &\{\luisK,\luisC\}=0  ,
&  \qquad  &
\endalignedat 
\tag6.6
$$
$$
\aligned
&  \Delta(\luisM)=1\otimes \luisM + \luisM\otimes 1 ,\\
&  \Delta(\luisH)=1\otimes \luisH +
\luisH\otimes (1 + \sigma \luisP)^2 ,\\
& \Delta(\luisD)=1\otimes \luisD + \luisD\otimes  \frac{1}{1 +
\sigma \luisP} - 
\frac{1}2 \luisM\otimes \frac{\sigma \luisP}{1 +
\sigma \luisP} ,\\
& \Delta(\luisC)=1\otimes \luisC + \luisC\otimes
\frac{1}{(1 + \sigma \luisP)^2} + \sigma   \luisD'\otimes
\frac{1}{1 + \sigma \luisP}\,\luisK \\
&\qquad\qquad \qquad + \frac{\sigma^2}{2}(\luisD')^2
\otimes
\frac{\luisM}{(1 +
\sigma \luisP)^2} ,\\
&  \Delta(\luisP)=1\otimes \luisP + \luisP\otimes 1 + \sigma
\luisP\otimes \luisP ,\\
&  \Delta(\luisK)=1\otimes
\luisK + \luisK\otimes  \frac{1}{1 + \sigma \luisP} +\sigma
\luisD'\otimes \frac{\luisM}{1 + \sigma \luisP} ,
\endaligned
\tag6.7
$$
where $\luisD'=\luisD+\frac 12 \luisM$.

 The Poisson-$gl(2)$ subalgebra generated by
$\{\luisM,\luisH,\luisD,\luisC\}$ is a  Schr\"odinger comodule algebra $V$ 
and the coaction 
$ 
\phi^{(2)}:gl(2)\rightarrow gl(2)\otimes {h_6^\sigma}
$ 
is given by the  restriction to the $gl(2)$ subalgebra of the full coproduct
map in ${h_6^\sigma}$~\cite{4}:
$$ 
  \phi^{(2)}(X):=\Delta (X)  ,\qquad 
 X\in \{\luisM,\luisH,\luisD,\luisC\}.
\tag6.8
$$ 
We take the following  symplectic realization  of ${h_6^\sigma}$: 
$$
\aligned
&  S(\luisC)=\frac{q_1^2}{2} ,\qquad
S(\luisH)=\frac{p_1^2}{2} ,\qquad
S(\luisD)=-p_1\,q_1 ,\\
& S(\luisM)=\lambda_1^2 ,\qquad\!\!\!
S(\luisK)=\lambda_1\,q_1 ,\quad\,
S(\luisP)=\lambda_1\,p_1 ,
\endaligned
\tag6.9
$$
and we consider a different symplectic realization  for the $gl(2)$ subalgebra
$$
T(\luisC)=\frac{q_1^2}{2} , \qquad
T(\luisH)=\frac{p_1^2}{2}+\frac{b_1}{q_1^2} ,\qquad
T(\luisD)=-p_1\,q_1 , \qquad T(\luisM)=\lambda_1^2  .
\tag6.10
$$
If we take as  Hamiltonian function 
$ 
H=\luisH + \luisC
$ 
we find that
$$ 
H^{(1)}_\sigma=T(H)=T(\luisH)+T(\luisC)
=\frac{p_1^2}{2}+\frac{q_1^2}{2}+\frac{b_1}{q_1^2},
\tag6.11
$$
 is just the undeformed $N=1$ SW Hamiltonian (3.4) with $\omega^2=1$.
But the  two-particle  case provides in a straightforward way a new integrable
deformation of the SW Hamiltonian  with comodule algebra symmetry:
$$
\aligned
& H^{(2)}_\sigma=(T\otimes S)(\phi^{(2)}(H))=(T\otimes
S)(\phi^{(2)}(\luisH) +\phi^{(2)}(\luisC)) \\
& \qquad =\frac{1}{2}(p_1^2 + p_2^2)+\frac{b_1}{q_1^2}
+ \frac{q_1^2}{2(1+\sigma\,\lambda_2\,p_2)^2}+\frac{q_2^2}{2} \\
& \qquad\qquad+ \sigma\,\lambda_2\left(2
\left(\frac{p_1^2}{2}+\frac{b_1}{q_1^2}\right)\,p_2 +
\frac{q_2(\lambda_1^2 - 2
q_1\,p_1)}{2(1+\sigma\,\lambda_2\,p_2)}\right) \\
&  \qquad\qquad\qquad +\sigma^2\,\lambda_2^2\,\left( 
\left(\frac{p_1^2}{2}+\frac{b_1}{q_1^2}\right)\,p_2^2 + 
\frac{(\lambda_1^2 - 2
q_1\,p_1)^2}{8(1+\sigma\,\lambda_2\,p_2)^2}   \right).
\endaligned
\tag6.12
$$
By considering the $gl(2)$  Casimir
function 
$
C_{V}=\frac{1}{4}\,\luisD^2 - \luisH\,\luisC,
$
the  constant of the motion  in involution with $ H^{(2)}_\sigma$ is obtained:
$$
C^{(2)}_\sigma=(T\otimes S)(\phi^{(2)}(C_{V}))=
(T\otimes S)(\frac{1}{4}\,\Delta(\luisD)^2 - \Delta(\luisH)) .
\tag6.13 
$$
As expected,  the limit $\sigma\rightarrow 0$ of $C^{(2)}_\sigma$ is just
$$
C^{(2)}_0=-\frac{1}{4}(p_2 q_1-p_1 q_2)^2 - \frac{b_1}{2}
\left(1+\frac{q_2^2}{q_1^2}\right).
$$
Further  iterations of the coaction map 
would  provide the corresponding integrable deformation in $N$ dimensions, but
in any case the only non-vanishing centrifugal term would be the one that
corresponds to $b_1$.

\bigskip

 We end with some remarks and open problems. Firstly, note that for {higher
rank coalgebras}, we have a set of right and left integrals coming from each of
the Casimir functions of the Poisson algebra. In general, these sets could not
be functionally independent  under an arbitrary symplectic realization and
the number of independent integrals coming from the coalgebra has to be fixed
for each individual realization. We also mention that subcoalgebras can also be 
used in order to extract superintegrability properties, as it was pointed out
in~\cite{2}. Finally, we think that the search for {explicit
solutions} of the deformed SW Hamiltonians and the corresponding deformed {Lax
formalism} are worthy to be considered in the future, as well as the
construction and analysis of the quantum mechanical analogues of  the deformed
SW Hamiltonians here introduced.


\bigskip
\bigskip
 \noindent
\head  Acknowledgments\endhead

\medskip
\noindent
 This work has been partially supported  by the Ministerio de Ciencia y
Tecnolog\1a, Spain (Project BFM2000-1055).
O.R. has been partially supported by INFN and by MIUR
(COFIN2001 ``Geometry and Integrability").  A.B. and F.J.H. are grateful to
CRM for hospitality.

\newpage


\bigskip
\bigskip

 \noindent
\head  References\endhead

 \medskip

 \eightpoint

\ref\key{1} 
\by  A. Ballesteros and F. J. Herranz
\paper   Integrable deformations of
oscillator chains from quantum algebras
\jour   J. Phys. A: Math. Gen.
\vol  32
\yr  1999
\pages  8851--8862
\endref

\ref\key{2} 
\by  A. Ballesteros and F. J. Herranz
\paper   Two-Photon Algebra and Integrable
Hamiltonian Systems
\jour  J. Nonlin. Math. Phys.  
\vol  8
\yr  2001
\pages  18--22
\endref

\ref\key{3} 
\by  A. Ballesteros, F. J. Herranz, J. Negro, and L.M. Nieto
\paper Twist maps for non-standard quantum algebras and discrete Schr\"odinger
symmetries  
\jour  J. Phys. A: Math. Gen.
\vol  33
\yr  2000
\pages  4859--4870
\endref

\ref\key{4} 
\by  A. Ballesteros, F. Musso, and O. Ragnisco
\paper  Comodule algebras and integrable
systems 
\jour  J. Phys. A: Math. Gen.
\vol  35
\yr  2002
\pages  8197--8211
\endref

\ref\key{5} 
\by  A. Ballesteros and O. Ragnisco
\paper   A systematic construction of
integrable Hamiltonians from coalgebras
\jour  J. Phys. A: Math. Gen.
\vol  31
\yr  1998
\pages  3791--3813
\endref

\ref\key{6} 
\by  A. Ballesteros and O. Ragnisco
\paper   Classical Hamiltonian systems
with  $sl(2)$ coalgebra symmetry and their integrable deformations
\jour  J. Math. Phys.
\vol  43
\yr  2002
\pages  954--969
\endref

\ref\key{7} 
\by  A. Ballesteros and O. Ragnisco
\paper  Coalgebra symmetry and superintegrability 
\jour  in preparation
\vol  
\yr  
\pages  
\endref

\ref\key{8} 
\by  G. Burdet, J. Patera, M. Perrin, and P. Winternitz
\paper   Optical group and its subgroups
\jour  J. Math. Phys.
\vol 19 
\yr  1978
\pages  1758--1780
\endref

\ref\key{9}
\by  V. Chari and A. Pressley       
\book  A Guide to Quantum Groups     
\publ  Cambridge University Press
\publaddr  Cambridge
\yr   1994      
\endref

\ref\key{10}
\by  E. E. Demidov, Y. I.  Manin, E. E. Mukhin, and 
 D. V. Zhdanovich
\paper   Nonstandard quantum deformations of $GL(N)$ and constants
solutions of the Yang--Baxter equation
\jour  Progr. Theor. Phys. Suppl.
\vol  102
\yr  1990
\pages  203--218
\endref

\ref\key{11}
\by  N. W.  Evans
\paper  Superintegrability in classical mechanics 
\jour  Phys. Rev. A
\vol  41
\yr  1990
\pages  5666--5676
\endref

\ref\key{12}
\by  N. W.  Evans
\paper   Superintegrability of the  Winternitz system
\jour  Phys. Lett. A
\vol  147
\yr  1990
\pages  483--486
\endref

\ref
\key{13}
\by   S. Majid      
\book  Foundations of Quantum Group Theory     
\publ  Cambridge University Press
\publaddr  Cambridge
\yr  1995       
\endref

\ref\key{14}
\by  F. Musso and O. Ragnisco
\paper Exact solution of the quantum
Calogero-Gaudin system and of its q-deformation  
\jour  J. Math. Phys.
\vol  41
\yr  2000
\pages  7386--7401
\endref

\ref\key{15}
\by  F. Musso and O. Ragnisco
\paper  The spin-1/2 Calogero-Gaudin system
and its $q$-deform\-ation 
\jour  J. Phys. A: Math. Gen.
\vol  34
\yr  2001
\pages  2625--2635
\endref

\ref\key{16}
\by  C.  Ohn
\paper   A star-product on $SL(2)$ and the corresponding
nonstandard quantum $U(SL(2))$
\jour  Lett. Math. Phys.
\vol  25
\yr  1992
\pages  85--88
\endref

\ref
\key{17}
\by  A. M. Perelomov       
\book   Integrable Systems of Classical
Mechanics and Lie algebras    
\publ  Birkh\"auser
\publaddr  Berlin
\yr      1990   
\endref

\ref\key{18}
\by   P. Winternitz, A. Smorodinsky, M. Uhlir, and J. Fris
\paper Symmetry Groups in Classical and Quantum Mechanics  
\jour Soviet J. Nuclear Phys. 
\vol 4 
\yr  1967
\pages  444
\endref

\ref\key{19}
\by   S. Zakrzewski
\paper  A hopf star-algebra of polynomials on the
quantum $SL(2,R)$ for a unitary $R$-matrix 
\jour Lett. Math. Phys. 
\vol  22
\yr  1991
\pages  287--289
\endref

\enddocument